\documentclass[reprint,floatfix]{revtex4-1}
\usepackage{amsmath,amssymb}
\usepackage{txfonts}
\usepackage{graphicx}
\usepackage{bm}
\usepackage{youngtab}
\usepackage{braket}
\newcommand{\vac}{\left|\mathrm{vac}\right\rangle} 

\usepackage{natbib}
\usepackage[pdfencoding=auto]{hyperref}

\usepackage[totalwidth=500pt,totalheight=680pt]{geometry}

\begin{document}

\title{Entanglement properties of the Haldane phases: A finite system-size approach}

\author{
Shohei Miyakoshi,$^{1}$
Satoshi Nishimoto,$^{2,3}$
and Yukinori Ohta$^{1}$
}
\affiliation{
$^{1}$Department of Physics, Chiba University, Chiba 263-8522, Japan\\
$^{2}$Institute for Theoretical Solid State Physics, IFW Dresden, 01171 Dresden, Germany\\
$^{3}$Department of Physics, Technical University Dresden, 01069 Dresden, Germany
}

\date{\today}

\begin{abstract}
We study the bond-alternating Heisenberg model using the finite-size density-matrix 
renormalization group (DMRG) technique and analytical arguments based on the matrix 
product state, where we pay particular attention to the boundary-condition dependence 
on the entanglement spectrum of the system.  
We show that, in the antiperiodic boundary condition (APBC), the parity quantum numbers 
are equivalent to the topological invariants characterizing the topological phases 
protected by the bond-centered inversion and $\pi$ rotation about $z$ axis.  
We also show that the odd parity in the APBC, which characterizes topologically nontrivial 
phases, can be extracted as a two-fold degeneracy in the entanglement spectrum even with 
finite system size.  
We then determine the phase diagram of the model with the uniaxial single-ion anisotropy 
using the level spectroscopy method in the DMRG technique.  
These results not only suggest the detectability of the symmetry protected topological (SPT) 
phases via general twisted boundary conditions but also provide a useful and precise numerical 
tool for discussing the SPT phases in the exact diagonalization and DMRG techniques.  
\end{abstract}


\maketitle

\section{Introduction}

Quantum spin models have long been studied in the field of strongly correlated electron systems.  
Since the discovery of the Haldane conjecture \cite{DHaldane1983A,DHaldane1983B}, qualitative
difference between systems of half-integer spins with gapless excitations and those of integer
spins with gapful excitations has attracted much attention.  In particular, the Affleck-Kennedy-Lieb-Tasaki 
model of spin $S=1$, which has a unique and analytically exact solution of the ground state,
was an important clue for quantum disordered phases of the integer spin systems
\cite{IAffleck1987,IAffleck1988}.  
The exact solution clarifies that exotic properties such as string orders and edge states
are observed in the Haldane phase and concomitantly that such quantum phases do not have
local order parameters.

According to the Landau-Ginzburg-Wilson (LGW) theory \cite{LLandau1958}, quantum phases
are classified by the spontaneous symmetry breakings and local order parameters.  
In this sense, the Haldane phase is a quantum disordered phase defined beyond the
LGW theory and often called the topological phase \cite{XGWen2004}. However, we do not yet
have a theoretical framework that enables one to identify the topological phases comprehensively.  
Generally, two gapful phases are identical if there is at least one path that connects the two phases
adiabatically without any spontaneous symmetry breaking or a gap closing.
In particular, the two phases that are distinguishable for a deformation under an imposed symmetry
are called the symmetry-protected topological (SPT) phases
\cite{XChen2011,XChen2012,FPollmann2010,FPollmann2012A,FPollmann2012B}.
The Haldane phase and topological insulators are known as examples of the SPT phase:
the former is a quantum phase that is protected by either the bond-centered inversion symmetry,
time reversal symmetry, or dihedral group ($\pi$ rotations about the $x$, $y$, and $z$ axes)
symmetry of the spin space \cite{FPollmann2010}, and the latter is a quantum phase protected
by the time reversal symmetry and U(1) gauge symmetry of charge \cite{CKane2005A,CKane2005B}.
Among the Haldane phases, the topologically nontrivial phase with even numbers of degenerate edge
states, which is called the odd-Haldane (OH) phase, is clearly distinguished from the topologically
trivial phase with odd numbers of degenerate edge states, which is called the even-Haldane (EH)
phase \cite{FPollmann2010,FPollmann2012A,FPollmann2012B,JKjall2013}.  

The Haldane phases have been studied by many analytical and numerical methods.  
In particular, the methods for classifying quantum disordered phases, such as the ones using
the string order parameters \cite{MNijs1989,MOshikawa1992,KTotsuka1995,MYamanaka1996},
quantized Berry phases \cite{THirano2008},
twisted order parameter \cite{MNakamura2002},
and level spectroscopy \cite{AKitazawa1997A,AKitazawa1997B,AKitazawa1997C,KNomura1998}, 
have achieved a great success.
The former string order parameters, which characterize a hidden $\mathbb{Z}_{2}\times\mathbb{Z}_{2}$
symmetry breaking in the Haldane phase, enable one to distinguish the Haldane phase with the
dihedral group symmetry \cite{MOshikawa1992}. The latter quantized Berry phase,
twisted order parameter, and level spectroscopy are often used in
finite-size systems, which are the methods using the difference in the quantum numbers
of the systems with twisted phases in an arbitrary bond and enable one to identify
the phase boundary clearly, unlike the methods of using the string order parameters.

Recently, a technique for distinguishing the SPT phases has been proposed \cite{HLi2008}, where the
entanglement spectrum (ES) is used.  The ES, which is the spectral structure appearing in the
reduced density matrix obtained by dividing the system into two subsystems, has much more information 
on the ground state than the entanglement entropy (EE) obtained as a von Neumann entropy of the 
reduced density matrix does.  It is known that the ES reproduces the spectral structure similar to the 
edge states in the SPT phase and that its two-fold degeneracy can be used as an index characterizing 
the difference between the trivial and nontrivial phases.  The ES is thus a powerful method for examining
the edge states such as the quantum Hall insulators, topological insulators, and other quantum spin chains.  
In the calculations of the ES, a variety of variational methods using the infinite matrix product
states (iMPS), such as infinite time-evolving block decimation (iTEBD) \cite{GVidal2007} and infinite
density-matrix renormalization group (iDMRG) methods \cite{IMcCulloch2008,JKjall2013}, as well as
the exact diagonalization and conventional DMRG methods
\cite{SWhite1992,SWhite1993A} for finite-size systems, have often been applied.

In this paper, motivated by the above developments in the field, we study the antiferromagnetic (AF) 
Heisenberg spin chains with a general spin quantum number $S$ and calculate the ES for finite-size 
systems of the model.  To find the degeneracy in the ES of the SPT phases generally requires 
sufficiently large subsystems.  
This is because the Schmidt decomposition keeping the two-fold degeneracy in the ES can only be 
achieved in the limit of large subsystems \cite{FPollmann2010}, as was confirmed by the direct 
calculation of the ES for the valence-bond-solid (VBS) wave function \cite{HKatsura2007,YXu2008}.  
Thus, the systems of large correlation lengths, such as those of a large spin $S$ or near critical points, 
the degeneracy of the ES is not exact, and therefore not necessarily an appropriate index characterizing 
the SPT phases.

We therefore study the boundary-condition and system-size dependences of the ES in the AF spin 
chains with periodic and antiperiodic boundary conditions based on the matrix product state (MPS).  
We show analytically that the parity quantum numbers in the antiperiodic boundary condition (APBC) 
are equivalent to the topological invariants in the SPT phases, which enables us to classify the phases.  
We also show that the parity quantum number leads to the two-fold degeneracy in the ES for the 
systems with the APBC and that the spin rotational symmetry leads to the quantization of the ES 
for the subsystem with a spin quantum number $S_A^z$.  
To confirm the validity of these proofs, we introduce the bond alternation $\delta$ to the model 
and study the behavior of the ES by numerical calculations using the DMRG technique.  

For systems with large $S$, where the gap decreases exponentially in the classical limit, the Haldane 
phase of the pure AF Heisenberg chain becomes unstable.  Moreover, the single-ion anisotropy 
$D$ leads to an instability of the in-plane AF ordering, which makes it difficult to determine the 
phase boundary due to the Berezinskii-Kosterlitz-Thouless (BKT) transition 
\cite{VBerezinskii1971,VBerezinskii1972,JKosterlitz1973}.  
We therefore apply the method of level spectroscopy using the PBC and APBC in the DMRG 
technique, with the help of the calculations of the central charge and string order parameter, and 
determine the phase boundaries for systems with $S=1, 2$, and $3$.  We in particular determine 
the phase diagram of the $S=2$ system in the parameter space of $\delta$ and $D$.  
We moreover find that the spin gap defined in the APBC reproduces not only the accurate 
gap-closing behavior but also the values of the Haldane gap in agreement with the previous 
numerical calculations.  The critical behavior at the transition points and topological properties of 
the system are also extracted by investigating the central charge and string order parameter.  
We thus clarify the entanglement properties of a variety of Haldane chains comprehensively.  

The rest of this paper is organized as follows.  
In Sec.~II, we define the bond-alternating Heisenberg model and discuss the methods of calculations 
used in this paper.  
In Sec.~III, we discuss the SPT phases of our model and construct the boundary condition by the 
MPS formalism.  We also clarify the meaning of the parity quantum number in the APBC.  
In Sec.~IV, we study the boundary and finite-size effects on the ES using the DMRG calculations.  
We also discuss the stability of the two-fold degeneracy in the ES from the viewpoint of the 
symmetry and corresponding quantum numbers.  
In Sec.~V, we discuss the effects of the single-ion anisotropy and determine the phase diagram of 
the $S=2$ model.  Summary of this paper is given in Sec.~VI.  

\section{Model and method}

\subsection{The model Hamiltonian}

Since the discovery of the Haldane phase, there are many analytical and
numerical studies of the Heisenberg AF (HAF) chains with integer spins
\cite{FPollmann2010,FPollmann2012A,FPollmann2012B,MNijs1989,MOshikawa1992,MYamanaka1996,
KTotsuka1995,THirano2008,MNakamura2002,AKitazawa1997A,AKitazawa1997B,AKitazawa1997C,
KNomura1998,JKjall2013,SWhite1993B,SEjima2015,STodo2001,HNakano2009,HKatsura2007,YXu2008,
MTsukano1998,RThomale2015,WChen2000,TTonegawa2011,USchollwock1995,USchollwock1996,KOkamoto2016}.
To study the entanglement properties of the Haldane phase, we consider the following AF chain 
with bond alternation $\delta$ and uniaxial single-ion anisotropy $D$ defined by the Hamiltonian
\begin{align}
\mathcal{H}=J\sum^{L}_{j=1}\bigl\{1+(-1)^{j}\delta\bigr\}
\bm{S}_{j}\cdot\bm{S}_{j+1}+D\sum_{j}\bigl(S^{z}_{j}\bigr)^{2} ,
\label{eq:BHAF}
\end{align}
where $J$ $(>0)$ is the AF exchange interaction (taken as a unit of energy) and
$\delta$ causes the dimerization transition.  $D$ $(>0)$ breaks the SU(2) symmetry
of the spin rotation, which leads to several gapful and gapless phases such as the
large-$D$ phase and in-plane AF phase.  However, neither of these terms breaks
any symmetry of the bond-centered inversion, time-reversal, and dihedral group
symmetry of spin space, which protect the Haldane phase.  

The qualitative picture of the HAF chain with bond-alternation can be obtained
from the $(1+1)$ dimensional O(3) nonlinear sigma model (NLSM) \cite{EFradkin2013},
which is derived from a semi-classical large-$S$ limit of the HAF chain.  
The O(3) NLSM is defined as follows:
\begin{align}
\mathcal{A}=\frac{v}{2g}\int\,d\tau dx\,
\Bigl\{\bigl(\partial_{x}\bm{n}\bigr)^{2}
+\frac{1}{v^{2}}\bigl(\partial_{\tau}\bm{n}\bigr)^{2}
\Bigl\}+i\Theta\,Q ,
\label{eq:O3NLSM}
\end{align}
where $g=2/S$, $v=2JS$, and the three-dimensional unit vector $\bm{n}(x)$ is
related to the spin operator as $\bm{S}_{j}/S\sim (-1)^{j}\bm{n}(x)+\bm{l}(x)$ \cite{EFradkin2013}.  
The term $i\Theta\,Q$ is called $\Theta$ term,
which is written as
\begin{align}
Q=\frac{1}{4\pi}\int d\tau dx
\bm{n}\cdot\partial\bm{n}
\times\partial\bm{n} .
\end{align}
Thus, $Q$ describes the integer-valued winding number.  If $\delta\ne 0$,
the $\Theta$ term is written as $\Theta=2\pi S(1+\delta)$.  
As proposed first by Haldane \cite{DHaldane1983A,DHaldane1983B}, the $\Theta$
term leads to the qualitative difference between the half-integer and integer spin systems.  
If $\Theta=0$, the O(3) NLSM model represents the gapful excitation, whereas
if $\Theta=\pi$, the model has the gapless excitation, which corresponds to the
massless free-boson theory that has the central charge $c=1$.  Therefore,
when the bond alternation changes from $-1$ to $1$, the phase
transitions with gap closing occur $2S$ times.

\subsection{Generalized valence-bond-solid state}

\begin{figure}
\centering
\includegraphics[width=0.6\columnwidth]{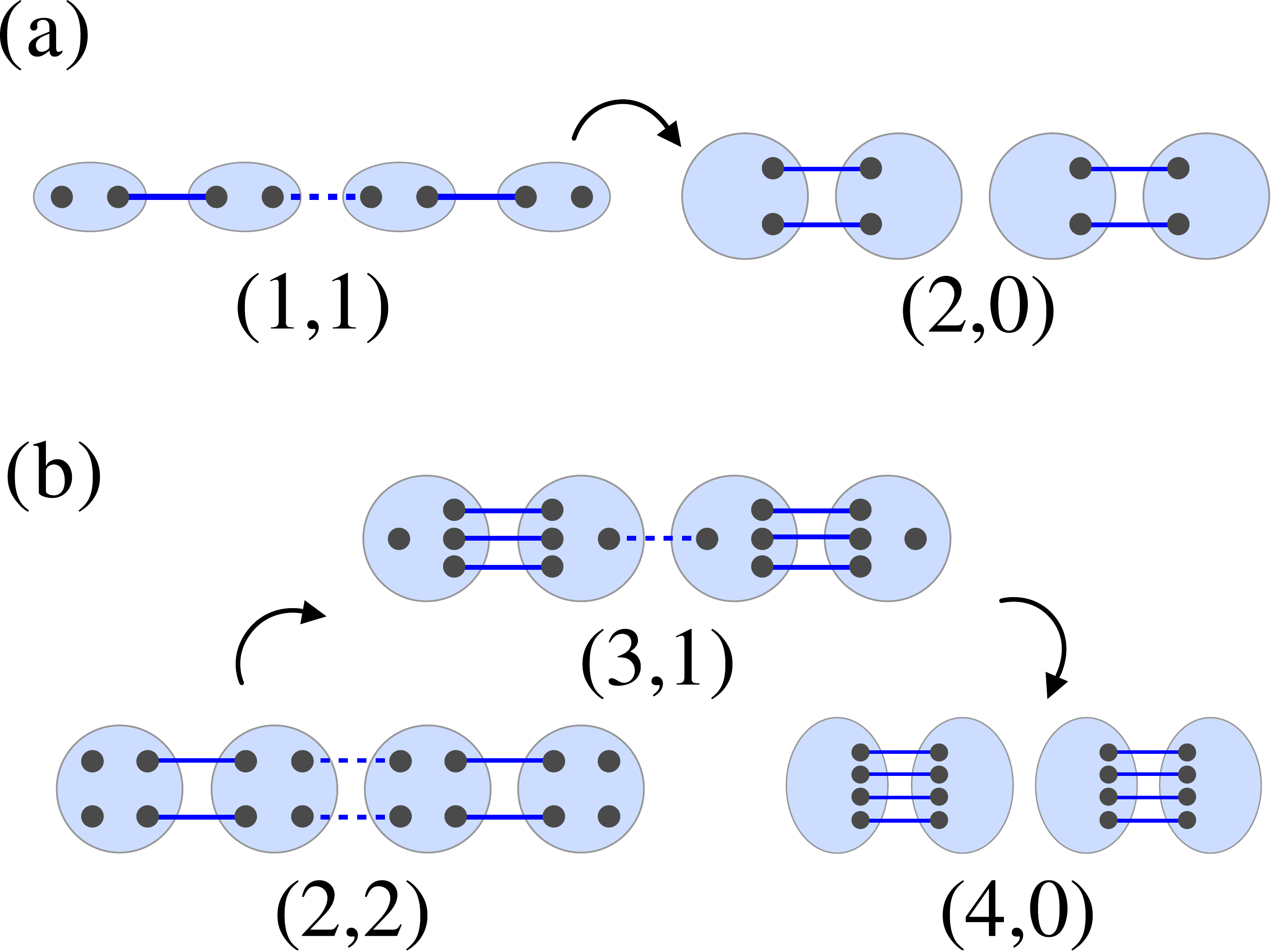}
\caption{
(Color online)
Schematic pictures of the dimerization transition in the bond-alternating HAF chains with
(a) $S=1$ and (b) $S=2$.  The links represent the spin-$1/2$ singlet bonds.  
The unconnected point represents the free edge spin.  
The panel (a) shows the Haldane-dimer transition, where we use the $(m,n)$-type VBS
state to represent the $(1,1)$ Haldane phase and $(2,0)$ dimer phase.  
The panel (b) shows the dimerization transition in the $S=2$ HAF chain with the $(2,2)$
Haldane phase, partially dimerized $(3,1)$ phase, and fully dimerized $(4,0)$ phase.  
}
\label{fig1}
\end{figure}

The phase transition in the bond-alternating system can be interpreted as the
change in the VBS configuration (see Fig.~\ref{fig1}).  
To see the qualitative properties of the gapped quantum phases, we introduce
the $(m,n)$-type generalized VBS state \cite{KTotsuka1995,MNakamura2002}
defined in the periodic boundary condition (PBC) as
\begin{align}
\ket{(m,n)}_\mathrm{PBC}=\frac{1}{\sqrt{\mathcal{N}}}
\prod^{L/2}_{j=1}\bigl(B^{\dagger}_{2j-1,2j}\bigr)^{m}
\bigl(B^{\dagger}_{2j,2j+1}\bigr)^{n}\vac
\label{eq:(m,n)-VBS} ,
\end{align}
where $B^{\dagger}_{i,j}=a^{\dagger}_{i}b^{\dagger}_{j}-b^{\dagger}_{i}a^{\dagger}_{j}$
with bosonic operators $a^{\dagger}_{i}$ and $b^{\dagger}_{i}$, $\mathcal{N}$ is
the normalization factor, and $\vac$ is the vacuum of bosons.  
Here, we use the Schwinger-boson representation of the spin operator defined as
$S^{+}_{i} = a^{\dagger}_{i}b_{i}$, $S^{-}_{i} = b^{\dagger}_{i}a_{i}$, and
$S^{z}=(a^{\dagger}_{i}a_{i}-b^{\dagger}_{i}b_{i})/2$.  
The integers $m$ and $n$ satisfy $m+n=2S$.  
If we consider the APBC defined as
$S^{z}_{L+1}=S^{z}_{1}$ and $S^{\pm}_{L+1}=e^{i\pi}S^{\pm}_{1}$
between $L$ and $1$ sites, the operator $B_{L,L+1}$ is written as
$B_{L,L+1}=i(a^{\dagger}_{L}b_{1}+b^{\dagger}_{L}a_{1})$.  
Thus, for the bond-centered inversion
$P:\, S_{i}\rightarrow S_{L-i+1}$,
the $(m,n)$-type VBS state has a parity quantum number defined as
\begin{align}
P\ket{(m,n)}_\mathrm{APBC}=
(-1)^{n+SL}\ket{(m,n)}_\mathrm{APBC}.
\label{eq:Inversion}
\end{align}
Therefore, the difference in the parity quantum number can be used
to identify the phase boundaries of the different VBS states.  
Moreover, the presence of these different phases is closely related to
the existence of the string order, which characterizes the breaking of the
hidden $\mathbb{Z}_{2}\times\mathbb{Z}_{2}$ symmetry \cite{MOshikawa1992}.  

\subsection{Level spectroscopy}

The level spectroscopy technique employing the APBC is a powerful tool for determining
the phase boundary between different VBS states.  
Due to the cancellation of the logarithmic corrections, this method can suppress the
finite-size effect and determine the phase diagram precisely.  
According to the previous studies of the level crossing
\cite{AKitazawa1997A,AKitazawa1997B,AKitazawa1997C,KNomura1998}, the gapful VBS 
phases and gapless in-plane AF phase can be characterized by the differences between 
the three lowest excitation energies defined as
\begin{align}
\Delta_\mathrm{EH}=&E_{0,\mathrm{APBC}}(0,+)-E_{0,\mathrm{PBC}}(0) , \\
\Delta_\mathrm{OH}=&E_{0,\mathrm{APBC}}(0,-)-E_{0,\mathrm{PBC}}(0) , \\
\Delta_{XY}=&E_{0,\mathrm{PBC}}(2)-E_{0,\mathrm{PBC}}(0) ,
\end{align}
where $E_{n,\mathrm{PBC}}(M)$ is the $n$-th lowest energy with the $z$-component
of the total spin $M=\sum_{j} S^{z}_{j}$ under the PBC and $E_{n,\mathrm{APBC}}(M,P)$ is
the $n$-th lowest energy with the $z$-component of the total spin $M$ and parity
quantum number $P$ under the APBC.  As was pointed out in the previous section, if
$\Delta_\mathrm{EH}$ ($\Delta_\mathrm{OH}$) is the lowest, the EH (OH) phase is
the most stable one.  
On the other hand, if the in-plane AF phase is the most stable state, the $\Delta_{XY}$ 
has the lowest energy.  We also use the spin gap defined as 
\begin{align}
\Delta_\mathrm{spin}=&
\left|\Delta_\mathrm{EH}-\Delta_\mathrm{OH}\right|
\label{eq:Spingap}
\end{align}
for evaluating the phase transition points between the EH and OH phases.  
The spin gap calculated by this definition is in good agreement with the results
of many previous numerical studies, such as the DMRG and quantum Monte Carlo (QMC)
calculations \cite{SEjima2015,STodo2001}, for the Haldane gap of the isotropic HAF chain because the
finite-size effect is rather small \cite{HNakano2009}.

\begin{figure}
\centering
\includegraphics[width=0.8\columnwidth]{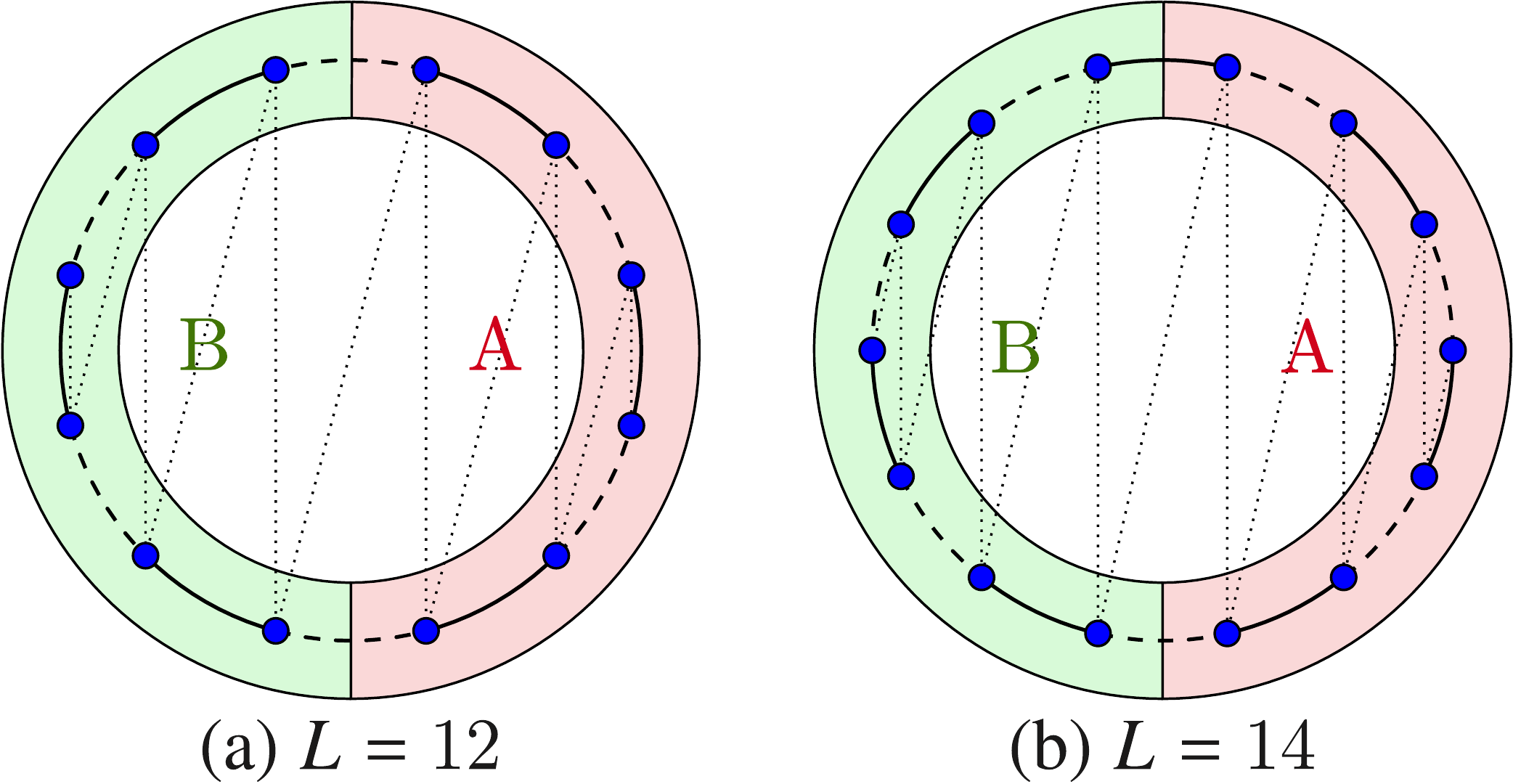} 
\caption{
(Color online)
Schematic pictures of the bipartition of the bond-alternating HAF chain used in the
entanglement spectrum calculation.  Alternating nearest-neighbor interactions
$J(1\pm\delta)$ are depicted by the solid and broken lines.  The red and green areas
indicate the subregions (or subsystems) $A$ and $B$, respectively.  
(a) The subregions are separated at the two strong (or weak) bonds that face each other.  
The system size is $L=12=0\,\,\mathrm{mod}\,\,4$.  
(b) The subregions are separated at the strong and weak bonds that face each other.  
The system size is $L=14=2\,\,\mathrm{mod}\,\,4$.  
}
\label{fig2}
\end{figure}

\subsection{Entanglement spectrum}

The entanglement related quantities have recently been studied extensively 
for investigating nonlocal correlations in many-body quantum states.  
In particular, Li and Haldane \cite{HLi2008} proposed that the ES is one of the powerful 
tools for investigating topologically ordered phases and symmetry protected (or enriched) 
topological phases, which are known as gapped phases with long-range and short-range 
entanglements, respectively.  
In our calculations, we consider the ES obtained by partitioning the system into two
subregions (or subsystems) $A$ and $B$ (see Fig.~\ref{fig2}).  Defining $\xi_{\lambda}$ in the 
Schmidt decomposition of the ground state $\ket{\psi}$ as
\begin{align}
\ket{\psi}=\sum_{\lambda}e^{-\xi_{\lambda}/2}
\ket{\lambda}_{A}\ket{\lambda}_{B} ,
\end{align}
where $\ket{\lambda}_{A}$ ($\ket{\lambda}_{B}$) is the orthonormal basis for the
subregion $A$ ($B$), we can interpret the ES as the energy spectrum of the
entanglement Hamiltonian $\mathcal{H}_{e}$ defined as
$e^{-\mathcal{H}_{e}}=\rho_{A}=\mathrm{Tr}_{B}\ket{\psi}\bra{\psi}
=\sum_{\lambda}e^{-\xi_{\lambda}}\ket{\lambda}_{A}\bra{\lambda}_{A}$.  
In particular, if the system size is sufficiently larger than the correlation length, the ES can
be described by the two virtual edge states; the ES therefore represents the gapless 
mode at real edges \cite{HLi2008}.  Indeed, it was rigorously proved that the degeneracy 
corresponding to the gapless edge mode can be found in the ES for 
the $(m,n)$-type VBS states \cite{HKatsura2007,YXu2008}.  
Figure~\ref{fig2} shows the examples of the bipartition of the bond-alternating Heisenberg
model in the periodic systems.  In the case of open boundary conditions, we need to
take into account the contribution from real edges or suitably chosen boundary constraints
for different VBS states.  To avoid these difficulties, we only consider the ES in the
periodic systems with the PBC and APBC.  For simplicity, we only discuss the bipartition
shown in Fig.~\,\ref{fig2}(b) in this paper, where the system is divided into two subregions
of length $L/2=\mathrm{odd}$.  Here, the $(m,n)$-type VBS state has $(m+1)(n+1)$ gapless
modes at real edges.  

The entanglement entropy (EE), which is defined as the von Neumann entropy
$S_{A}=-\mathrm{Tr}\bigl[\rho_{A}\ln{\rho_{A}}\bigr]$, also has important information
on the topological phases and criticality of the system.  
From the conformal field theory \cite{PCalabrese2004}, the EE in the periodic systems
can be obtained as
\begin{align}
S_{A}(l)=\frac{c}{3}\ln{\Biggl[\frac{L}{\pi}
\sin{\Biggl(\frac{\pi l}{L}\Biggr)}\Biggr]}+s_1 ,
\end{align}
where $s_{1}$ is a nonuniversal constant and $l$ is the length of the partition A.  
$c$ is the central charge characterizing the critical theory describing the low-energy
physics.  To determine the central charge, we evaluate the terms \cite{SNishimoto2011}
\begin{align}
c^{*}(L)\equiv\frac{3[S_{A}(L/2-d)-S_{A}(L/2)]}{\ln{[\cos{(\pi d/L)}]}}.
\label{eq:Centralcharge}
\end{align}
from the direct calculations of $S_{A}(l)$ assuming the PBC.  
$d$ is the lattice constant.  The values of $c^{*}(L)$ converge to the central
charge $c$ in the infinite system-size limit.  

\section{Parity in the antiperiodic boundary condition}

As discussed in Sec.~II B, the $(m,n)$-type VBS state in the APBC has a parity 
quantum number with respect to the bond inversion operation.  
The difference in the parity quantum numbers, which is applied to the level spectroscopy, 
has so far played a major role in the numerical determination of the boundary 
between the VBS states.  
In this section, we consider the meaning of the parity quantum numbers in the APBC 
based only on the MPS formalism and symmetry arguments. 
First, using the MPS, we clarify the meaning of the SPT phases that are derived under 
the assumptions of the translational symmetry of the bond alternating chain and the 
bond-centered inversion symmetry of the lattice.  
Next, we clarify the properties of the operators corresponding to the APBC by 
constructing the MPS from the exact VBS wave functions.  
Also, based on the above discussions, we extend our theory to the general MPS and 
clarify the relationship between the boundary conditions and parity quantum numbers.  
Moreover, we clarify the equivalence between the parity quantum numbers and topological 
invariants on the basis of the classification of the SPT phases by the $\pi$ rotation about 
$z$ axis and bond-centered inversion operation.  

\subsection{Matrix-product-state formalism}

First, let us discuss the classification of the SPT 
phases in the bond-alternating systems.  
In the one-dimensional system, the MPS is a good approximation for the gapped ground 
state \cite{GVidal2007,DPerezGarcia2007}.  
Thus, the MPS formalism can rigorously prove the presence of several SPT phases.
Here, we introduce the classification of
the SPT phases by the bond-centered inversion symmetry and translational symmetry 
of the dimerized lattice.  First, we define the MPS as
\begin{align}
\ket{\psi}=&\sum_{i_{1},i_{2},\cdots,i_{L}}\mathrm{Tr}
\Bigl[\Lambda^{A}\Gamma^{A}_{i_{1}}
\Lambda^{B}\Gamma^{B}_{i_{2}}\cdots
\Lambda^{A}\Gamma^{A}_{i_{2N}}\Bigr]
\nonumber\\
&\times\ket{i_{1},i_{2},\cdots,i_{2N}} ,
\end{align}
where $\Lambda^{a}$ $(a=A, B)$ is a $\chi_{a}\times\chi_{a}$ positive matrix, and
$\Gamma^{A}$ and $\Gamma^{B}$ are $\chi_{A}\times\chi_{B}$ and $\chi_{B}\times\chi_{A}$
matrices, respectively.  Here, we define $N=L/2$.  $i_{n}$ represents the 
physical degrees of freedom of site $n$.  
The MPS representation is not unique for given states, but we can choose the
canonical MPS \cite{DPerezGarcia2007} satisfying $\mathrm{Tr}\Bigl[(\Lambda^{A})^{2}\Bigr]=
\mathrm{Tr}\Bigl[(\Lambda^{B})^{2}\Bigr]=1$ and
\begin{align}
&\sum_{m}\Gamma_m^{A}\Lambda^{B}\Lambda^{B}(\Gamma_m^{A})^{\dagger}=
\sum_{m}(\Gamma_m^{B})^{\dagger}\Lambda^{A}\Lambda^{A}\Gamma_m^{B}=
\mathbb{I}_{\chi_{B}} , \label{eq:canonical1}\\
&\sum_{m}\Gamma_m^{B}\Lambda^{A}\Lambda^{A}(\Gamma_m^{B})^{\dagger}=
\sum_{m}(\Gamma_m^{A})^{\dagger}\Lambda^{B}\Lambda^{B}\Gamma_m^{A}=
\mathbb{I}_{\chi_{A}} , \label{eq:canonical2} 
\end{align}
where $\mathbb{I}_{\chi_{a}}$ is a $\chi_{a}\times\chi_{a}$ matrix.  The canonical
conditions Eqs.~(\ref{eq:canonical1}) and (\ref{eq:canonical2}) imply that the transfer
matrix has a left (right) eigenvector $\mathbb{I}_{\chi_{A}}$ ($\mathbb{I}_{\chi_{B}}$) with
eigenvalue $\lambda=1$.  Moreover, since the MPS must be a pure state, we assume
that $\mathbb{I}_{\chi_{a}}$ is the only eigenvector with the largest eigenvalue $1$.  

Next, let us consider the inversion symmetry at a bond between the sites $1$ and $2N$.  
Since the MPS is invariant under the bond-centered inversion, there exists
a unitary transformation $U^{a}_P$ with $[U^{a}_P,\Lambda^{a}]=0$ such that
\begin{align}
&\bigl(\Gamma^{A}_{m}\bigr)^{T}=e^{i\theta^{A}_P}
\bigl(U^{B}_P)^{\dagger}
\Gamma^{B}_{m}U^{A}_P ,
\label{eq:InversionA}
\\
&\bigl(\Gamma^{B}_{m}\bigr)^{T}=e^{i\theta^{B}_P}
\bigl(U^{A}_P)^{\dagger}
\Gamma^{A}_{m}U^{B}_P ,
\label{eq:InversionB}
\end{align}
where $\theta^{a}_P$ is a phase.  Using the above relation twice, we obtain
\begin{align}
\sum_{m}\Gamma^{A}_{m}\Lambda^{B}
U^{B}_P(U^{B}_P)^{*}
\Lambda^{B}(\Gamma^{A}_{m})^{\dagger}
&=e^{-i(\theta^{A}_P+\theta^{B}_P)}
U^{A}_P(U^{A}_P)^{*} ,
\\
\sum_{m}\Gamma^{B}_{m}\Lambda^{A}
U^{A}_P(U^{A}_P)^{*}
\Lambda^{A}(\Gamma^{B}_{m})^{\dagger}
&=e^{-i(\theta^{A}_P+\theta^{B}_P)}
U^{B}_P(U^{B}_P)^{*} ,
\end{align}
where we use the canonical conditions Eqs.~(\ref{eq:canonical1}) and (\ref{eq:canonical2}).  
Moreover, due to the assumption of the pure MPS, we obtain
$U^{a}_P(U^{a}_P)^{*}=e^{i\phi^{a}_P}\mathbb{I}_{\chi_{a}}$
and $2(\theta^{A}_P+\theta^{B}_P)=0\,\mathrm{mod}\,2\pi$.  
Using the canonical condition and these results, we obtain the following relation:
\begin{align}
\theta^{A}_P+\theta^{B}_P
-\phi^{A}_P+\phi^{B}_P=0\,\mathrm{mod}\,2\pi.
\end{align}
As a consequence, by the bond-centered inversion symmetry, we can distinguish four
different states
$(\theta^{A}_P+\theta^{B}_P,\phi^{A}_P,\phi^{B}_P)=(0,0,0)$, $(0,\pi,\pi)$, $(\pi,\pi,0)$, and $(\pi,0,\pi)$,
which are separated by the quantum phase transitions.  
The large-$D$ phase $\ket{D}=\ket{0}\ket{0}\cdots\ket{0}$ is a trivial phase, where
the matrices $\Gamma^{a}$, $\Lambda^{a}$, and $U^{a}_P$ are scalars.  
Thus, the large-$D$ phase belongs to the state with
$(\theta^{A}_P+\theta^{B}_P,\phi^{A}_P,\phi^{B}_P)=(0,0,0)$.

The other three states can be described as the generalized VBS states.  Using the
MPS, the $(m,n)$-type VBS state of Eq.~(\ref{eq:(m,n)-VBS}) can be rewritten as
\begin{align}
\ket{(m,n)}_\mathrm{PBC}=&
\sum_{i_{1},i_{2},\cdots,i_{2N}}
\mathrm{Tr}\Bigl[g^{A}_{i_{1}}
g^{B}_{i_{2}}\cdots g^{B}_{i_{2N}}\Bigr]
\nonumber\\
&\times\ket{i_{1},i_{2},\cdots,i_{2N}},
\end{align}
where the matrices are given by
\begin{align}
\bigl\{g^{A}_{l}\bigr\}_{q,p}&=
(-1)^{n-q}\sqrt{{}_{n-q+p}C_{p}\,{}_{m-p+q}C_{q}}~\delta_{l=(n-m)/2-q+p} , 
\\
\bigl\{g^{B}_{l'}\bigr\}_{p,q}&=
(-1)^{m-p}\sqrt{{}_{m-p+q}C_{q}\,{}_{n-q+p}C_{p}}~\delta_{l'=(m-n)/2-p+q}
\end{align}
with $0\le p\le m$ and $0\le q\le n$.  
$l$ and $l'$ are indexes of the local quantum states
satisfying $-(m+n)/2\le l$ and $l'\le (m+n)/2$, respectively. 
We ignore the normalization factor that does
not affect our discussion.  If we introduce the matrices
$\bigl\{u^{A}\bigr\}_{qq'}=(-1)^{q}\delta_{q+q'=n}$ and
$\bigl\{u^{B}\bigr\}_{pp'}=(-1)^{p}\delta_{p+p'=m}$, where $0\le p$, $p'\le m$ and
$0\le q,q'\le n$, we obtain the following relations:
\begin{align}
\bigl\{(u^{B})^{\dagger}g^{B}_{i}u^{A}\bigr\}_{p,q}
&=(-1)^{m}(g^{A}_{i})^{T}_{p,q} , 
\nonumber\\
\bigl\{(u^{A})^{\dagger}g^{A}_{i}u^{B}\bigr\}_{q,p}
&=(-1)^{n}(g^{B}_{i})^{T}_{q,p} .
\end{align}
These results correspond to Eqs.~(\ref{eq:InversionA}) and (\ref{eq:InversionB}) with
$e^{i(\theta^{A}_P+\theta^{B}_P)}=(-1)^{n+m}=(-1)^{2S}$.  
Therefore, the $(m,n)$-type VBS states with integer spins describe the states with
$(\theta^{A}_P+\theta^{B}_P,\phi^{A}_P,\phi^{B}_P)=(0,0,0)$
for even $m$ and $n$ and $(0,\pi,\pi)$ for odd $m$ and $n$.  
The $(m,n)$-type VBS states with half-integer spins describe the states with
$(\theta^{A}_P+\theta^{B}_P ,\phi^{A}_P,\phi^{B}_P)=(\pi,\pi,0)$
and $(\pi,0,\pi)$ for even $n$ and odd $n$, respectively.  
Moreover, if the system recovers the translational symmetry of the 
original lattice without dimerization, the SPT phases with 
$(\theta^{A}_P+\theta^{B}_P,\phi^{A}_P,\phi^{B}_P)=(0,0,0)$
and $(0,\pi,\pi)$ are connected to the so-called even Haldane (EH) 
and odd Haldane (OH) phases, respectively.  Note that the SPT phases with
$(\theta^{A}_P+\theta^{B}_P ,\phi^{A}_P,\phi^{B}_P) =(\pi,\pi,0)$ and $(\pi,0,\pi)$ vanish.
In the above discussions, we assume the translational symmetry of the dimerized lattice.    
However, even if there is only the inversion symmetry about a bond center, the classification 
of the SPT phases can be discussed for systems with any larger unit cells 
\cite{FPollmann2010}.  
The change in the translational symmetry due to lattice dimerization thus plays an 
important role for classifying the phases in the $(m,n)$-type VBS state.  

\subsection{Meanings of the parity quantum number}

Next, let us consider the parity quantum number in the APBC.  Here, we present a general symmetry 
argument without using the exact VBS wave functions, which is based only on the MPS formalism.  
To treat the boundary condition, we now rewrite the $(m,n)$-type VBS state in the MPS 
representation as follows: 
\begin{align}
\ket{(m,n)}_\mathrm{APBC}=&
\sum_{i_{1},i_{2},\cdots,i_{2N}}
\mathrm{Tr}\Bigl[
U^{A}_\mathrm{tw}g^{A}_{i_{1}}g^{B}_{i_{2}}
\cdots g^{B}_{i_{2N}}\Bigr]
\nonumber\\
&\times\ket{i_{1},i_{2},\cdots,i_{2N}} ,
\end{align}
where we define $\{U^{A}_\mathrm{tw}\}_{q,q'}=(-1)^{n-q}\delta_{q,q'}$.  
As in the Schwinger-boson argument for the bond-centered inversion, 
the above MPS representation can also reproduce Eq.~(\ref{eq:Inversion}).  
Here, we note that the matrix $U^{A}_\mathrm{tw}$ satisfies 
$u^{A}(U^{A}_\mathrm{tw})^{T}(u^{A})^{\dagger}=(-1)^{n}U^{A}_\mathrm{tw}$.  
Thus, this argument on the APBC can be generalized as follows: 
If there is a unitary matrix $U^{A}_\mathrm{tw}$ corresponding to  
the projective representation of some symmetry satisfying 
\begin{align}
U^{A}_P(U^{A}_\mathrm{tw})^{T}(U^{A}_P)^{\dagger}
=e^{i\phi^{A}_\mathrm{tw}}U^{A}_\mathrm{tw},
\label{eq:Parity}
\end{align}
where $e^{i\phi^{A}_\mathrm{tw}}$ is a topological invariant characterizing the SPT phases,
then the MPS with a boundary condition defined as
\begin{align}
\ket{\psi}=&\sum_{i_{1},i_{2},\cdots,i_{2N}}
\mathrm{Tr}\Bigl[U^{A}_\mathrm{tw}\Lambda^{A}\Gamma^{A}_{i_{1}}
\Lambda^{B}\Gamma^{B}_{i_{2}}\cdots\Lambda^{B}\Gamma^{B}_{i_{2N}}
\Bigr]\nonumber\\&\times
\ket{i_{1},i_{2},\cdots,i_{2N}}
\label{eq:Twisted}
\end{align}
should have the parity quantum number as the topological invariant 
$e^{i\phi^{A}_\mathrm{tw}}$. 
In particular, as long as the global $\pi$ rotation about the $z$ axis, 
$R_{z}$, is not broken, we can reproduce the APBC by choosing 
$U^{A}_\mathrm{tw}$ to be the projective representation of $R_{z}$.  
Here, the parity of the MPS in Eq.~(\ref{eq:Twisted}) can be written as
$e^{-i(\phi^{A}_{z,P}-\phi^{A}_{z})}$.  Details of the proof are given in Appendix A.
The projective representation of $P$ itself satisfies Eq.~(\ref{eq:Twisted}),
so that $e^{-i\phi^{A}_{P}}$ corresponds to the parity.  
The twisted boundary condition can also be applied to the cases with other symmetries
such as dihedral group and time-reversal symmetries \cite{XChen2015}.  In the case of the
dihedral group of the spin rotation, we find that the spin reversal operation gives us similar
topological invariants.

\section{Entanglement spectrum in the bond-alternating Heisenberg model}

In this section, we discuss our numerical DMRG results for the entanglement properties of 
the bond-alternating HAF chains with $S=1, 2$, and $3$.
First, we determine the phase boundaries and then calculate the spin gap under the APBC, 
which signals the transition between the EH and OH phases.  
Secondly, we evaluate the ES of the $(m,n)$-type VBS state and their symmetry protected 
properties in the periodic systems.  Here and hereafter, we pay particular attention to the $S=2$ case.  
Thirdly, we discuss the stability of the two-fold degeneracy in the ES from the viewpoint of 
the symmetries and corresponding quantum numbers.  
Moreover, making the finite-size scaling analysis of the ES, we clarify the equivalence between 
the two boundary conditions PBC and APBC and the edge-ES correspondence in the 
thermodynamic limit.  

\subsection{Phase boundaries}

\begin{figure*}[htbp]
\centering
\includegraphics[width=1.8\columnwidth]{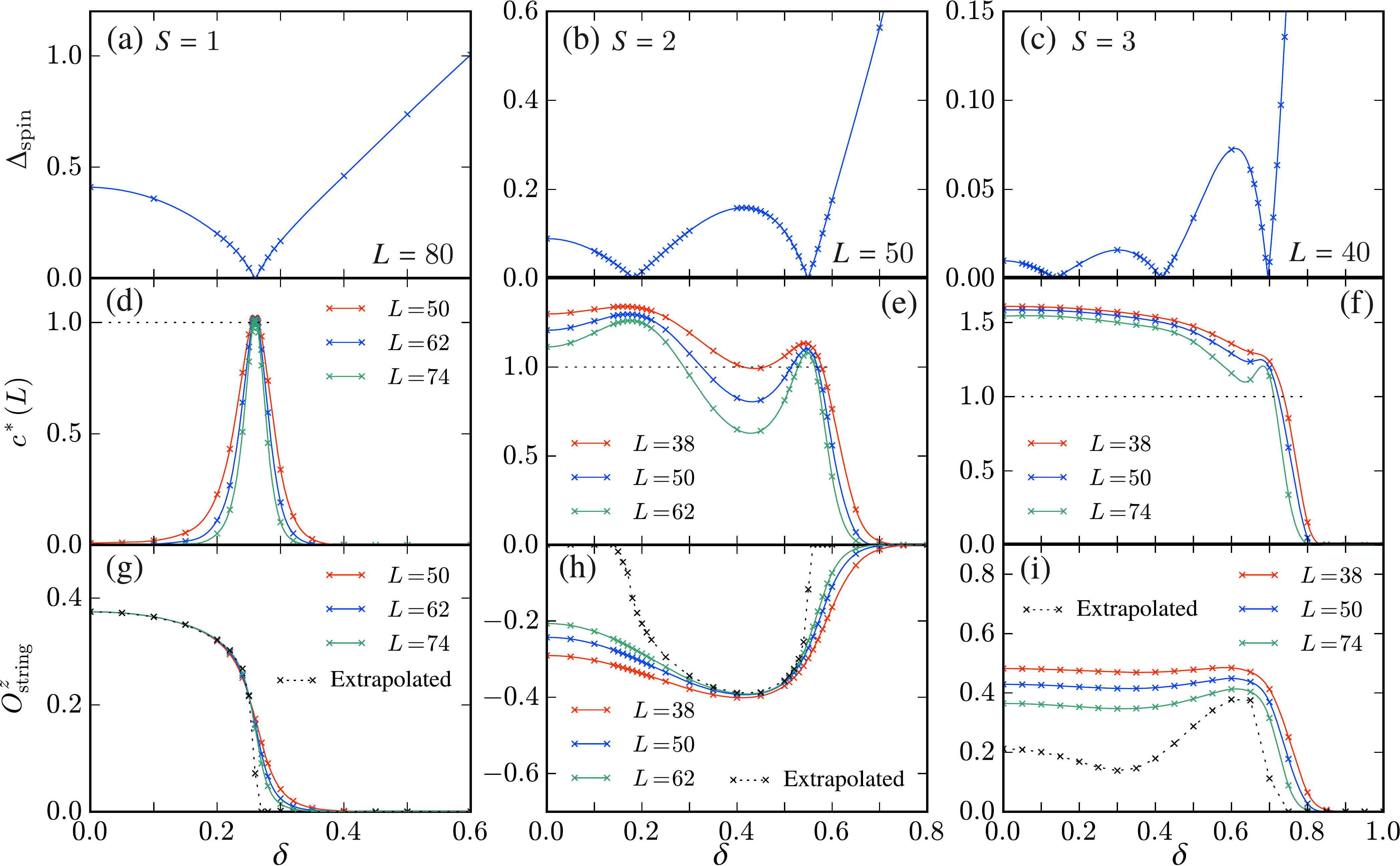}
\caption{
(Color online)
Calculated quantities as a function of $\delta$ in the bond-alternating HAF chain
Eq.~(\ref{eq:BHAF}) for $S=1$, $2$, and $3$.  
Upper panels (a)-(c):  
spin gap $\Delta_\mathrm{spin}$ calculated using Eq.~(\ref{eq:Spingap}) with
$L=80$ for $S=1$, with $L=50$ for $S=2$, and with $L=40$ for $S=3$.  
Middle panels (d)-(f):
central charge $c^{*}(L)$ calculated using Eq.~(\ref{eq:Centralcharge}).  
Lower panels (g)-(i):
string order parameter $O^{z} _\mathrm{string}$ calculated using Eq.~(\ref{eq:Stringorder}).  
Here, the dotted line shows the size extrapolation adopting Eq.~(\ref{eq:Extrapolation}).  
In the calculations of the central charge and string order parameter, we use the PBC
with $L$ up to $74$ for $S=1$, with $L$ up to $62$ for $S=2$, and with $L$ up to $74$ for $S=3$.  
The peak positions of the central charge clearly correspond to the level crossing points
between the EH and OH phases in the APBC.  
}
\label{fig3}
\end{figure*}

First, let us discuss the phase boundaries of the bond-alternating HAF chains with 
$S=1$, $2$, and $3$.  To determine the phase boundaries between the different VBS states, 
we evaluate the three quantities: 
spin gap $\Delta_\mathrm{spin}$, 
string order parameter $O^{z}_\mathrm{string}$, 
and central charge $c^{*}(L)$.  
The calculated results are shown in Fig.~\ref{fig3}.  

We note here that the qualitative behaviors of the bond-alternating HAF chain can be obtained from
the O(3) NLSM; in particular, this model at the phase transition point is described by the SU(2)
symmetric Tomonaga-Luttinger liquid, which corresponds to the conformal field theory with the
central charge $c=1$.  
As shown in Fig.~\ref{fig3}(d), our numerical result for $S=1$ is in quantitative agreement with this
argument.  Moreover, the transition point obtained from the spin gap precisely coincides with the
peak positions of the result for the central charge $c^{*}(L)$.  
From the level-crossing point of $S=1$ with $L=80$, we obtain $\delta=0.25995J$, which is in precise
agreement with the previous result of the QMC calculations \cite{MNakamura2002}, $\delta_{c}=0.25997(3)J$.  
Similarly, for $S=2$ and $3$, the transition points are also in good agreement with the results of the
previous studies \cite{MNakamura2002}.  The details of the transition points for $S=1/2,1,\cdots,4$ are
summarized in Appendix B.  
We moreover find that the $S=1$ Haldane gap obtained without size extrapolation at $\delta=0$
becomes $\Delta_{S=1}=0.401479J$, which is again in precise agreement with the QMC result
$\Delta_{S=1}=0.41048(6)J$ \cite{STodo2001} as well as a recent DMRG result
$\Delta_{S=1}=0.41047924(4)J$ \cite{SEjima2015}.  
Similarly, we obtain the Haldane gaps for $S=2$ and $3$ as $\Delta_{S=2}=0.088653J$ and
$\Delta_{S=3}=0.009763J$, respectively, which are also in quantitative agreement with the previous 
QMC results \cite{STodo2001}, $\Delta_{S=2}=0.08917(4)J$ and $\Delta_{S=3}=0.01002(3)J$.  
We note that the central charges for $S=2$ and $3$ show peak-like structures at the level crossing
points (see Fig.~\ref{fig3}).  However, the data for $c^{*}(L)$ at the peaks do not have the value
$c=1$ precisely, which may be caused by the large correlation lengths for $S=2$ and $3$ systems.  

The lower panels of Fig.~\ref{fig3} show the results for the string order parameter 
$O^{z}_\mathrm{string}$, which is defined as
\begin{align}
O^{z}_\mathrm{string}(\left|j-k\right|=L/2)=
\left\langle
S^{z}_{j}\exp{\Bigl(i\pi\sum^{k-1}_{l=j}S^{z}_{l}\Bigr)}S^{z}_{k}
\right\rangle.
\label{eq:Stringorder}
\end{align}
The nonzero string order parameter can be used to distinguish between the topologically
trivial EH phase and nontrivial OH phase as long as the dihedral group symmetry remains.  
This definition, however, depends on the positions of the strong and weak bonds.
Thus, we use the system with $L=4n+2$ $(n\in\mathbb{Z})$ sites and consider the case
with $j=\,\mathrm{even}$ and $k=\,\mathrm{odd}$ in the PBC (see Fig.~\ref{fig2}).  
The string order parameter thus calculated shows a very slow convergence with respect
to $L$ due to the large correlation length in the $S\ge 2$ systems.  In the extrapolations,
we assume the correlation function at a distance $L/2$ to be of the form
\begin{align}
O^{z}_\mathrm{string}(L/2)\sim
A+B\exp{\bigl(-L/\lambda\bigr)}/L^{\gamma} ,
\label{eq:Extrapolation}
\end{align}
where $\lambda$ and $\gamma$ are positive constants.  
Here, we use a least-square fitting
with $L=38,50,62,74$ for $S=1$,
with $L=26,38,50,62$ for $S=2$, and
with $L=38,50,62,74$ for $S=3$.  
We confirm that the extrapolated values for $S=1$ and $2$ are in good agreement with the
transition points obtained from the spin gap.  

\begin{figure*}[htbp]
\centering
\includegraphics[width=2.0\columnwidth]{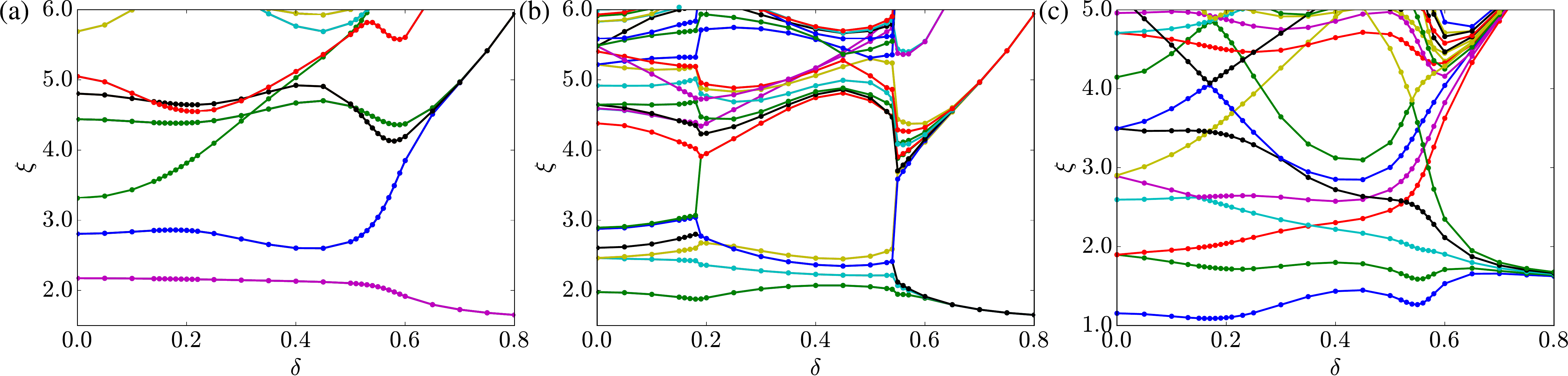}
\caption{(Color online)
Calculated entanglement spectra of the $S=2$ bond-alternating HAF chain with $L=50$ as a
function of $\delta$ under the (a) PBC and (b) APBC.  Panel (c) shows the results for
$L=50$ in the APBC obtained by applying the staggered magnetic field $h_{z}/J=0.001$.  
Note that the staggered magnetic field breaks the bond-centered inversion and other
symmetries that protect the Haldane phase.  Thereby, (c) shows an adiabatic continuation
of the entanglement spectra, which signals nonexistence of the SPT phases.  
}\label{fig4}
\end{figure*}

\subsection{Entanglement spectrum of the dimer Haldane phase}

Next, let us discuss the ES for the ground state of the bond-alternating HAF chain.  
The gapless modes for the real edges become in general a good explanation of the
degeneracies in the ES.  Indeed, using the $(m,n)$-type VBS state, the correspondence
between the degeneracies in the ES and those in the edge modes for the real edges 
was identified analytically \cite{HKatsura2007,YXu2008}.  
In the periodic systems, the $(m,n)$-type VBS state has an $(n+1)(m+1)$-fold
[$(n+1)(n+1)$-fold or $(m+1)(m+1)$-fold] degeneracy for the bipartition shown
in Fig.~\ref{fig2}(b) [Fig.~\ref{fig2}(a)].  
For simplicity, we consider the ES for the bipartition shown in Fig.~\ref{fig2}(b).  
Moreover, we consider two types of the periodic boundary conditions, APBC and PBC,
to study the ES of the finite-size systems.  Figures \ref{fig4}(a) and \ref{fig4}(b)
show calculated results for the ES of the $S=2$ bond-alternating HAF chain for $L=50$
in the PBC and APBC, respectively, where the spectra are normalized as
$\sum_{\lambda}e^{-\xi_{\lambda}}=1$.    
We find that the ES shows different behaviors, depending on the boundary conditions:
In the PBC, the ES deforms continuously as the bond alternation $\delta$ increases.    
In contrast, the ES in the APBC shows sudden changes at $\delta = 0.18$ and $0.55$,
which correspond, respectively, to the level crossing between the $(2,2)$ and $(3,1)$
phases and between the $(3,1)$ and $(4,0)$ phases.  Note that all the spectra in the
$(3,1)$ phase are doubly degenerate but such degeneracy does not appear in the $(2,2)$ 
and $(4,0)$ phases.  From the viewpoint of the level degeneracy, we identify that the larger
contributions to the ES ($e^{-\xi_\lambda}>10^{-3} \sim 10^{-4}$) shown in Figs.~\ref{fig4}(a)
and \ref{fig4}(b) are roughly consistent with the ES of the VBS ground state in the
infinite system, where the $(n,m)$-state has the $(n+1)(m+1)$-fold degeneracy in
the ES.  It should be noted that the ES in the PBC does not show a complete two-fold
degeneracy due to the finite-size effect.  
As was rigorously proved in \cite{HKatsura2007,YXu2008}, the two-fold degeneracy
in the ES is preserved as long as the system size is sufficiently larger than the
correlation length \cite{SEjima2015}.  
In the APBC, there is some freedom in selecting the bipartition shown in Fig.~\ref{fig2}
and in selecting the bond of the phase $\pi$ twisting;  however, we find that the result
depends only on the geometric difference in the bipartition but does not depend on
the bond of the phase $\pi$ twisting.  

Let us consider the staggered magnetic field $\sum_{j}(-1)^{j}h_{z}S^{z}_{j}$ here, which
breaks the inversion symmetry about the bond center and other symmetries protecting
the Haldane phase.  Figure \ref{fig4}(c) shows the ES in the APBC with $h_{z}=0.001J$.  
We find that the ES deforms continuously as the bond alternation increases, i.e., no
level crossings, suggesting that, under the staggered magnetic field, the Haldane phase
does not exist or the topological invariant $e^{-i(\phi^{A}_{z,P}-\phi^{A}_{z})}=\pm 1$
disappears.  This absence of the Haldane phase in the staggered magnetic field is 
in agreement with the result of Ref.~\onlinecite{MTsukano1998}.

\begin{figure*}[htbp]
\centering
\includegraphics[width=1.3\columnwidth]{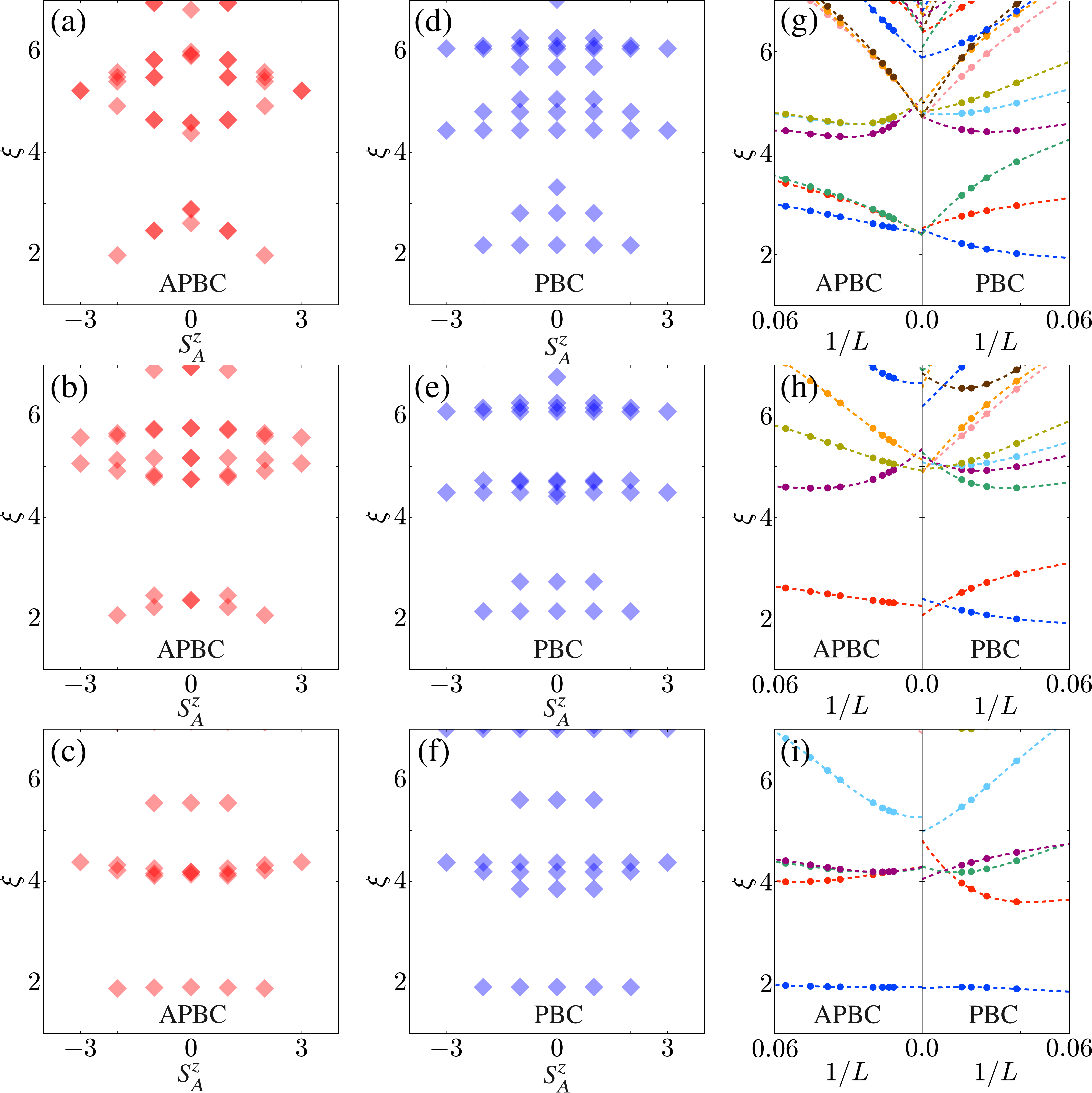}
\caption{
(Color online)
Entanglement spectra of the model with $S=2$ as the function of the $z$-component of
the total spin of the subregion A, $S^{z}_{A}$, calculated using the APBC (left panels) and
PBC (middle panels) with $L=50$.  
Upper, middle, and lower three panels are for $\delta=0$, $0.4$, and $0.6$, respectively.  
Right panels (g)-(i) show the finite-size extrapolations of the entanglement spectra for the
PBC and APBC with $S^{z}_{A}=0$.  
}
\label{fig5}
\end{figure*}

\subsection{Degeneracy in the entanglement spectrum}

Now, let us discuss details of the degeneracy in the ES.  Here, we first show analytically the
degenerate structure of the ES from the viewpoint of the symmetry.  We consider the case
of the PBC first.  Here, we pay attention to the following properties of the ground state:
(i) the system has the SU(2) symmetry and
(ii) the ground state has a total spin quantum number $S_\mathrm{tot}=0$.  
Hence, the ground state can be written as a state, in which the total angular momentum 
$S_\mathrm{tot}$ of the subregions A and B is zero:
\begin{align}
\ket{\psi}=\sum_{s}
a_{s}\ket{S_\mathrm{tot}=0,M=0;S_{A}=s,S_{B}=s},
\end{align}
where $S_\mathrm{tot}$, $M$, $S_A$, and $S_B$ are the quantum numbers of
$\bm{S}^{2}_\mathrm{tot}$, $S^{z}_\mathrm{tot}$, $\bm{S}^{2}_{A}$,
and $\bm{S}^{2}_{B}$, respectively.
Noting the Clebsch-Gordan coefficients to be
$\braket{s,m,s,-m|0,0;s,s}=(-1)^{s-m}/\sqrt{2s+1}$,
we can rewrite the ground state as
\begin{align}
\ket{\psi}=\sum_{s}\sum^{s}_{m=-s}
\frac{(-1)^{s-m}}{\sqrt{2s+1}}
a_{s}\ket{s,m}_{A}\ket{s,-m}_{B}.
\end{align}
The reduced density matrix $\rho_{A}$ is then written as
\begin{align}
\rho_{A}=\sum_{s}\frac{a^2_{s}}{2s+1}
\sum^{s}_{m=-s}\ket{s,m}_{A}\bra{s,m}_{A} ,
\end{align}
which leads to the result that the degeneracy in the PBC depends only on the quantum
number of the subregion A, $\bm{S}^{2}_{A}$. This result also holds when the system
has an open boundary condition. However, when the ground state of the system is
ferromagnetic, this result does not necessarily hold because of the condition (ii).  
In fact, the ES behaves differently depending on $S^{z}_{A}$ \cite{RThomale2015}.  

Next, we consider the case of the APBC. The above result cannot be used because the
APBC breaks the SU(2) symmetry.  Here, the parity about the bond-centered inversion
gives an important contribution to the two-fold degeneracy in the ES.  
Assuming the bond-centered inversion symmetry $P$ between equally divided
subregions $A$ and $B$ of the system, we can choose the wave function satisfying
\begin{align}
P\ket{i}_{A}=\ket{i+1}_{B},~~
P\ket{i+1}_{B}=\ket{i}_{A},
\end{align}
where $\ket{i}_{A}$ ($\ket{i}_{B}$) is the orthonormal basis in the subregion $A$ ($B$)
and $i$ ($=\textrm{even}$) is an index of the wave function.  Thereby, all the wave functions
in this system can be written as
\begin{align}
\ket{\psi}=\sum_{i=\textrm{even}}\sum_{j=\textrm{odd}}
c_{i,j}\ket{i}_{A}\ket{j}_{B} ,
\end{align}
where $c_{i,j}$ is a complex coefficient satisfying the normalization condition
$\sum_{i,j}|c_{i,j}|^{2}=1$.  If we apply the bond-centered inversion $P$ on the
above wave function, we obtain the relation
\begin{align}
P\ket{\psi}=&\sum_{i=\textrm{even}}\sum_{j=\textrm{odd}}
c_{i,j}\ket{j-1}_{A}\ket{i+1}_{B}
\nonumber\\
=&\sum_{i=\textrm{even}}\sum_{j=\textrm{odd}}
c_{j-1,i+1}\ket{i}_{A}\ket{j}_{B}.
\end{align}
Thus, the wave function with a parity $\eta=\pm$ has the coefficients $c_{i,j}$ satisfying
$c_{i,j}=\eta c_{j-1,i+1}$.  Let us then introduce a complex matrix $a_{i,j}=c_{2i,2j+1}$, so that
$a_{i,j}=\eta a_{j,i}$.  Using this relation, we can write the wave function as
\begin{align}
\ket{\psi}=\sum_{i,j=0}a_{i,j}
\ket{2i}_{A}\ket{2j+1}_{B} .
\end{align}
When the parity is odd, the matrix $a_{i,j}$ is written as a complex skew-symmetric matrix.  
Generally, a complex skew-symmetric matrix can be written as a block diagonalized form
containing $e^{i\phi}i\sigma^{y}$ on a proper basis, where we use the Pauli matrix $\sigma^{y}$.  
Thus, the above wave function can be written as
\begin{align}
\ket{\psi}=\sum_{k}\lambda_{k}e^{i\phi_{k}}
\Bigl(\ket{k,1}_{A}\ket{k,2}_{B}-\ket{k,2}_{A}\ket{k,1}_{B}\Bigr) ,
\end{align}
where $\ket{k,1}_{A(B)}$ and $\ket{k,2}_{A(B)}$ are the orthonormal bases and $\lambda_{k}$ 
is a real coefficient satisfying $2\sum_{k}\lambda_{k}^{2}=1$.
The reduced density matrix can then be obtained as 
\begin{align}
\rho_{A}=\mathrm{Tr}_{B}\ket{\psi}\bra{\psi}=\sum_{k}\lambda_{k}^{2}
\Bigl(\ket{k,1}_{A}\bra{k,1}_{A}+\ket{k,2}_{A}\bra{k,2}_{A}\Bigr) . 
\end{align}
Thereby, we find that the reduced density matrix for the odd parity phase has the two-fold
degeneracy.  We therefore find that, due to the odd parity, the OH phase in the APBC can be
detected as the two-fold degeneracy in the ES.  
Note that, in the PBC, the VBS states do not have an odd parity even in the nontrivial cases, 
and thus the ES does not show a stable two-fold degeneracy in any finite-size systems.  
We also note that the direct calculation for the $(m,n)$-type VBS state in the APBC leads 
to the confirmation of the presence of the two-fold degeneracy in the ES.  

Now, to confirm the validity of the proofs given above, we consider the $S^z_{A}$ dependence
of the ES calculated numerically.  Our numerical results for the ES in the PBC and APBC
are shown in Figs.~\ref{fig5} (a)-(f), which are given as a function of $S^{z}_{A}$ at $L=50$, 
where $S^{z}_{A}$ is the $z$-component of the total spin of the subregion A.  
In the APBC, we find that all the spectra at $\delta=0.4$ are doubly degenerate but that a
part of the spectra at $\delta=0$ and $0.6$ does not show the degeneracy.  This difference
in the degenerate structure comes from the difference in the parity quantum number, whereby
we find that the phase at $\delta=0.4$ belongs to the $(3,1)$ VBS state, whereas the phases
at $\delta=0$ and $0.6$, which have the even parity, belong to the $(2,2)$ VBS and $(4,0)$
VBS states, respectively.  These results are consistent with the results shown in Fig.~\ref{fig3}.  
In the PBC, in contrast, $S_{A}$ becomes a good quantum number in all the parameter space.  
This is due to the SU(2) symmetry of the system and the spin quantum number
$S_{\rm tot}=0$ of the wave function.  

Looking, in particular, at the low-lying states of the ES in the APBC and PBC shown in
Fig.~\ref{fig5}, we find that the same numbers of the degenerate spectra, which are separated by
the gaps, are present.  If we consider that the low-lying ES consists of two free edge spins 
with $S=n/2$ and $m/2$, the degeneracy becomes $(n+1)(m+1)$-fold, which is consistent with
the numbers of the spectra observed in Figs.~\ref{fig5}(a)-(f).  The result in the PBC is also
consistent with the fusion rule of the SU(2) symmetry: i.e.,
$3\otimes3=1\oplus3\oplus5$, $4\otimes2=3\oplus5$, and $5\otimes1=5$.  
Here, the lack of the exact $(n+1)(m+1)$-fold degeneracy in the PBC is interpreted to be due
to the interference between the edge spins because the system size is smaller than the
correlation length.  Therefore, the ES in the PBC does not show the two-fold degeneracy
even in the $(3,1)$ VBS state, which is in contrast to the exact two-fold degeneracy in the
ES observed in the APBC.

To clarify the points given above further, let us make the size extrapolation of the ES.  
Figures \ref{fig5}(g)-(i) show the results for $\delta=0$, $0.4$, and $0.6$ with $S^{z}_{A}=0$
in the APBC and PBC, where we assume the polynomial function of form
$\xi(L)=A+B/L+C/L^{2}+D/L^{3}$ for the extrapolation.  We thus find that the degeneracy in
the ES in the APBC and PBC are approximately in agreement with each other at
$L\rightarrow\infty$.  
This result not only indicates that the difference in the boundary conditions can be
neglected in the thermodynamic limit but also suggests that the wave functions in the
APBC and PBC become locally equivalent to each other in the thermodynamic limit.  
In other words, the absence of the SU(2) symmetry in the APBC cannot be seen 
in the ES in the thermodynamic limit, whereas in the PBC, the difference in the 
topological triviality or nontriviality appears in the degeneracy of the ES in this limit.  
We thus find that the two-fold degeneracy in the ES established using the MPS formalism 
\cite{FPollmann2010} is consistent with the results of the size extrapolations of 
finite-size calculations with different boundary conditions.  

\begin{figure}[htbp]
\centering
\includegraphics[width=\columnwidth]{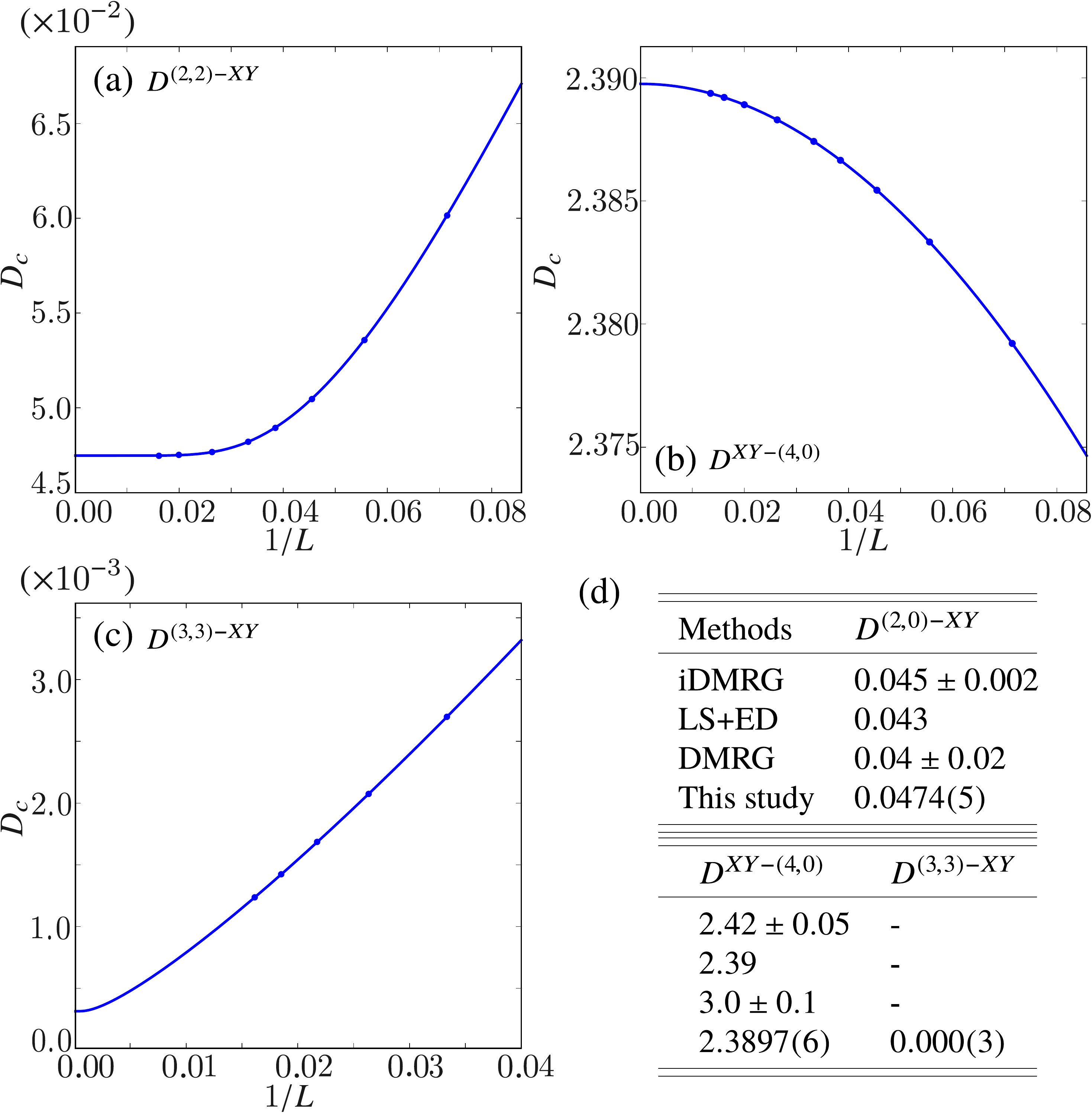}
\caption{(Color online)
System-size dependence of the BKT transition points $D_{c}$ of the $S=2,3$ HAF chains
with the uniaxial single-ion anisotropy.  
(a) [(c)] Transition point between the $(2,2)$ [$(3,3)$] VBS state and in-plane AF phase (XY).  
(b) Transition point between the in-plane AF phase and $(4,0)$ VBS state as the large-$D$ phase.  
The size-extrapolated values of our results are shown in (d).  
The iDMRG, LS+ED (level spectroscopy plus exact diagonalization), and DMRG results are 
taken from Refs.~\onlinecite{JKjall2013,USchollwock1995,USchollwock1996,KNomura1998}.
}
\label{fig6}
\end{figure}

\begin{figure}[htbp]
\centering
\includegraphics[width=0.7\columnwidth]{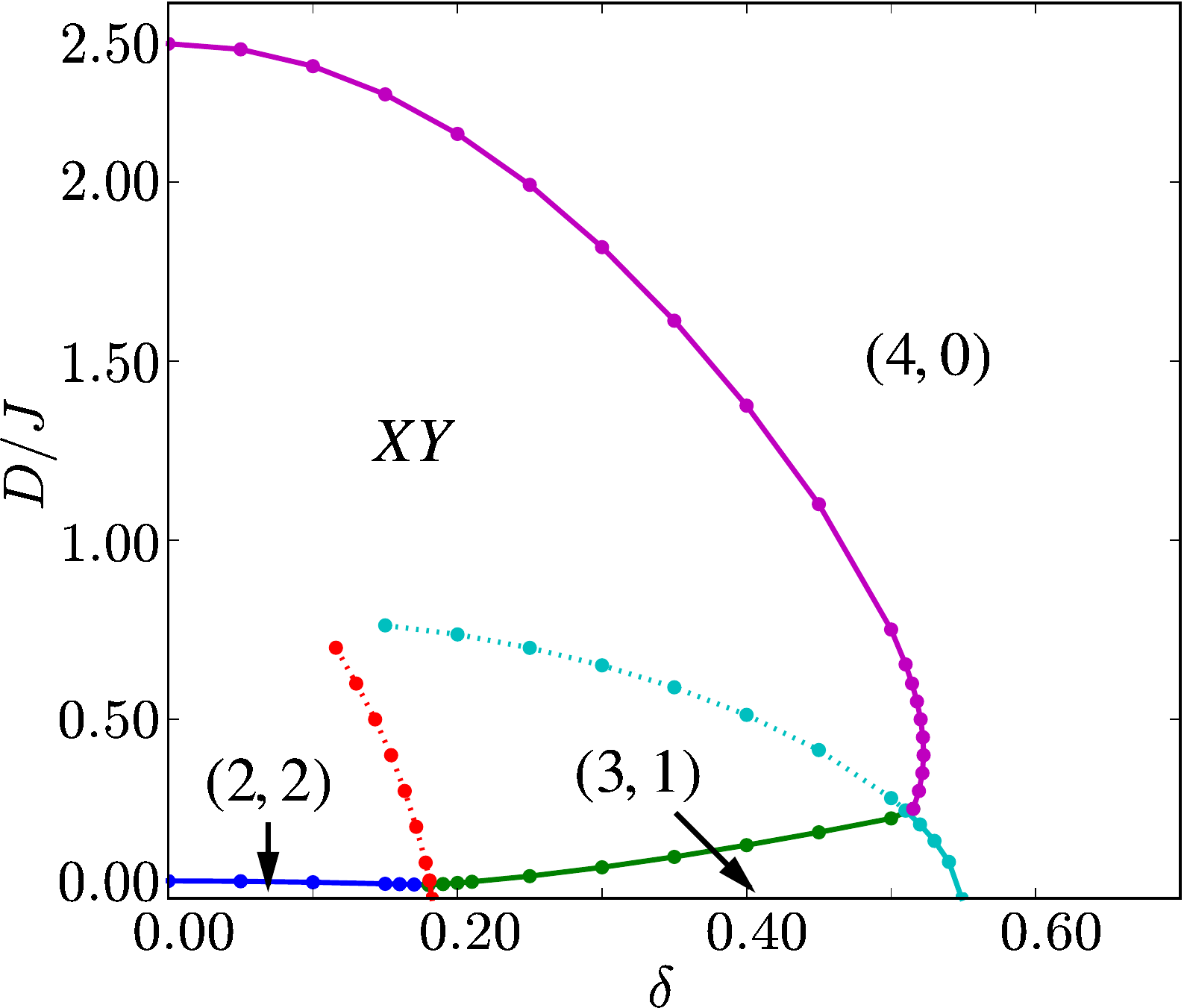}
\caption{(Color online)
Calculated phase diagram of the $S=2$ HAF chain in the parameter space of the 
bond alternation $\delta$ and uniaxial single-ion anisotropy $D$.  The phase boundaries 
between the $(2,2)$, $(3,1)$, and $(4,0)$ VBS states are determined by the level crossing
of the excitation energies, $\Delta_\mathrm{EH}$ and $\Delta_\mathrm{OH}$.  
The region of the in-plane AF phase ($XY$) is determined as the region where the 
excitation energy $\Delta_{XY}$ is the lowest among the three energies 
$\Delta_\mathrm{EH}$, $\Delta_\mathrm{OH}$, and $\Delta_{XY}$.  
}
\label{fig7}
\end{figure}

\section{Uniaxial single-ion anisotropy}

Finally, let us discuss the effect of the uniaxial single-ion anisotropy $D$ in the
bond-alternating HAF chain.  The uniaxial single-ion anisotropy $D$ $(>0)$ causes
the topologically trivial large-$D$ phase and gapless in-plane AF phase ($XY$)
in the HAF without bond alternation.  Therefore, the HAF chain with this anisotropy
leads to the Gaussian transition between the different VBS states and also to
the BKT transition between the gapful and gapless phases.  
In the HAF chain with $S=2$, in particular, this anisotropy is conjectured to lead
to the gapful intermediate phase called the intermediate-$D$ phase \cite{MOshikawa1992}.  
However, the existence of these phases has not sufficiently been worked out,
except for systems with special terms such as $D_{4}\sum_{j}(S^{z}_{j})^{4}$.  
This is in particular the case when the system shows the BKT transition with quite
a large correlation length.  There are, however, several methods for approaching
the BKT transition.  As discussed in the previous section, to use the level
spectroscopy obtained from the conformal field theory gives us one of the powerful
solutions.  Moreover, recent methods using iMPS, which treats infinite-size systems
within a reasonable truncation error, are expected to solve this problem \cite{JKjall2013}.  

Here, we first present the results for the phase boundaries of the BKT transition
in the HAF chains with $S=2$ and $3$ (see Fig.~\ref{fig6}).  We use the level
spectroscopy technique in the DMRG for large-size systems.  
In the previous calculations using iDMRG, the BKT transition points for
$(2,2)-XY$ and $XY-(4,0)$ are obtained as $D_{c}/J=0.045\pm0.002$ and
$D_{c}/J=2.42\pm0.05$, respectively.  Our results, on the other hand, show a saturation
behavior as a function of $L$ as shown in Fig.~\ref{fig6}(a), whereby we can determine
the phase transition points for $(2,2)-XY$ and $XY-(4,0)$ as $D_{c}/J=0.0474(5)$ and
$D_{c}/J=2.3897(6)$, respectively, in a high precision.  
Thus, the level spectroscopy technique using the DMRG offers a very accurate
method for determining the BKT transition points.  We note, however, that the transition
point for $(3,3)-XY$ is hard to determine even with our technique, which is due to
quite a large correlation length.  

Next, let us discuss the phase diagram of the model for $S=2$ in the parameter space
of $\delta$ and $D$.  As was discussed in Ref.~\onlinecite{WChen2000} for $S=1$, the $(2,0)$
dimer phase and large-$D$ phase are adiabatically connected with each other, but
the gapless in-plane AF phase does not appear in the $S=1$ case.  In Fig.~\ref{fig7},
we show the phase diagram in the $S=2$ case, which is calculated using the level
spectroscopy for $L=26$ in the DMRG technique.  As seen in Figs.~\ref{fig6}(a) and \ref{fig6}(b),
we find that a qualitatively correct phase diagram can be obtained without size extrapolation,
which is little changed even in the thermodynamic limit.
Actually, the $S=2$ phase diagram in a wide parameter space of $\delta$ and $D$ was 
obtained recently by an exact diagonalization and level spectroscopy analysis of 
small clusters \cite{KOkamoto2016}.
As in the case of $S=1$, we find that the large-$D$ phase is connected with the $(4,0)$
dimerized phase.  However, in the case of $S=2$, the phase diagram is mostly covered
by the in-plane AF phase.  This behavior is also found in the phase diagram in the
parameter space of the uniaxial single-ion anisotropy and Ising anisotropy
\cite{TTonegawa2011,JKjall2013}, and may be due to the large correlation length in the
large-$S$ systems; in other words, this result reflects the classical-spin--like behavior
of the HAF chain.  

\section{Summary}

In this paper, we have studied the entanglement properties of the bond-alternating HAF chain.  
We determined the phase boundaries of the different VBS states with high precision via the 
DMRG method employing the level-spectroscopy technique.  We found that the spin gap 
defined in the APBC reproduces not only the accurate gap-closing behavior but also 
the values of the Haldane gap in agreement with the previous numerical calculations.  
Moreover, investigating the central charge and string order parameter, we extracted the 
critical behavior at the transition points and topological properties of the system.  
We however found that for $S>1$ the proper size extrapolation is required to determine 
the accurate phase boundaries due to the large correlation length.  We also discussed the 
effect of the uniaxial single-ion anisotropy in the bond-alternating HAF chain.  
We confirmed the adiabatic continuation between the large-$D$ phase and $(4,0)$ dimer 
phase for $S=2$, which is in qualitative agreement with the $S=1$ case.  We also found 
that the phase diagram for $S=2$ is covered largely by the in-plane AF phase.  

We have moreover studied the ES in the APBC using not only the DMRG calculation but 
also the symmetry argument from the MPS formalism.  Considering the boundary effects 
on the degeneracy of the ES in finite systems, we elucidated the following: 
First, we analytically proved the equivalence between the parity quantum numbers in the 
APBC and the topological invariants, which enables us to classify the SPT phases by use 
of the $\pi$ rotation about $z$ axis and bond-centered inversion operation.  
Secondly, we showed that the odd parity in the APBC, which characterizes the topologically 
nontrivial phase, leads to the two-fold degeneracy in the ES.  
Thirdly, evaluating the ES in the thermodynamic limit by the DMRG method, we confirmed 
that the ES in the PBC and APBC complementarily recovers the edge-state picture of 
the low-lying ES.  
These arguments obtained in this paper are based on the general MPS formalism.  Thus, 
we can apply this method to the classification of other SPT phases with general boundary 
conditions and also to the evaluation of their entanglement quantities.  In particular, our 
theory, if extended to general twisted boundary conditions, will not only provide a useful 
numerical tool for investigating other SPT phases but also offer valuable clues for 
identifying novel SPT phases.

\begin{acknowledgments}
We thank M.~Nakamura for enlightening discussions.  
This work was supported in part by a Grant-in-Aid for Scientific Research (No.~26400349) 
from JSPS of Japan.  S.~M.~acknowledges support from the JSPS Research Fellowship 
for Young Scientists and hospitality of IFW Dresden during his stay in Dresden and his use 
of computers.  S.~N.~acknowledges financial support from the German Research Foundation 
(Deutsche Forschungsgemeinschaft, DFG -- SFB-1143).  
\end{acknowledgments}

\begin{appendix}

\section{Detailed proof for the parity quantum number}

To make our discussion on the parity quantum number in the APBC complete,
we here consider the SPT classification for the bond-centered inversion $P$
and $\pi$ rotation about $z$ axis $R_{z}$.  The projective representation of $R_{z}$ 
may be defined as
\begin{align}
R_{z}\Gamma^{A}=e^{i\theta^{A}_{z}}
(U^{A}_{z})^{\dagger}\Gamma^{A} U^{B}_{z},
\label{eq:rotateA}
\\
R_{z}\Gamma^{B}=e^{i\theta^{B}_{z}}
(U^{B}_{z})^{\dagger}\Gamma^{B} U^{A}_{z},
\label{eq:rotateB}
\end{align}
where we ignore the suffixes $m$ for simplicity.  Using the above relation
twice, we obtain $(U^{a}_{z})^{2} = e^{i\phi^{a}_{z}}\mathbb{I}_{\chi_{a}}$.  
Then, considering the case where the two operators $R_{z}$ and $P$ act,
we obtain the following relations using Eqs.~(\ref{eq:InversionA}) and (\ref{eq:rotateA}):
\begin{align}
&PR_{z}\Gamma^{A}=e^{i(\theta^{A}_{z}+\theta^{A}_{P})}(U^{B}_{z})^{T}(U^{B}_{P})^{\dagger}\Gamma^{B}U^{A}_{P}(U^{A}_{z})^{*},
\\
&R_{z}P\Gamma^{A}=e^{i(\theta^{B}_{z}+\theta^{A}_{P})}(U^{B}_{P})^{\dagger}(U^{B}_{z})^{\dagger}\Gamma^{B}U^{A}_{z}U^{A}_{P}.
\end{align}
Using the relation $PR_{z}\Gamma^{A}=R_{z}P\Gamma^{A}$, we obtain
\begin{align}
e^{i(\theta^{A}_{z}-\theta^{B}_{z})}\Gamma^{B}=~ & U^{B}_{P}(U^{B}_{z})^{*}(U^{B}_{P})^{\dagger}(U^{B}_{z})^{\dagger}
\nonumber\\
& \times\Gamma^{B}U^{A}_{z}U^{A}_{P}(U^{A}_{z})^{T}(U^{A}_{P})^{\dagger} .
\end{align}
Similarly, exchanging the suffixes $A$ and $B$, we obtain
\begin{align}
e^{i(\theta^{B}_{z}-\theta^{A}_{z})}\Gamma^{A}=~ & U^{A}_{P}(U^{A}_{z})^{*}(U^{A}_{P})^{\dagger}(U^{A}_{z})^{\dagger}
\nonumber\\ & \times
\Gamma^{A}U^{B}_{z}U^{B}_{P}(U^{B}_{z})^{T}(U^{B}_{P})^{\dagger} .
\end{align}
Since $U^{A}_{P}(U^{A}_{z})^{*}(U^{A}_{P})^{\dagger}(U^{A}_{z})^{\dagger}$ becomes
an eigenvector of the transfer matrix, we find
\begin{align}
U^{A}_{z}U^{A}_{P}=e^{i\phi^{A}_{z,P}}U^{A}_{P}(U^{A}_{z})^{*} .
\end{align}
Noting the relations  $U^{A}_{P}=e^{i\phi^{A}_{P}}(U^{A}_{P})^{T}$ and $U^{A}_{z}=e^{i\phi^{A}_{z}}(U^{A}_{z})^{\dagger}$,
we find
\begin{align}
U^{A}_{z}U^{A}_{P}=e^{i(\phi^{A}_{z,P}+\phi^{A}_{P}-\phi^{A}_{z})}(U^{A}_{z}U^{A}_{P})^{T} ,
\end{align}
where $U^{A}_{z}U^{A}_{P}$ corresponds to the projective representation of the combined symmetry $R_{z}P$.  
Therefore, we obtain the relation
\begin{align}
\phi^{A}_{z,P}+\phi^{A}_{P}-\phi^{A}_{z}=0, \pi\,\mathrm{mod}\,2\pi,
\end{align}
which leads to further classification of the SPT phases \cite{FPollmann2010}.  
Moreover, because the matrix $U^{z}_{P}$ satisfies the relation
\begin{align}
U^{A}_{P}(U^{A}_{z})^{T}(U^{A}_{P})^{\dagger}=e^{-i(\phi^{A}_{z,P}-\phi^{A}_{z})}U^{A}_{z} ,  
\end{align}
we find that the parity quantum number in the APBC can be written as the topological invariant
$e^{-i(\phi^{A}_{z,P}-\phi^{A}_{z})}=\pm 1$.  In particular, we find $\phi^{A}_{z,P}-\phi^{A}_{z}=\pi$
in the OH phase, so that the OH phase has an odd parity, resulting in the two-fold degeneracy 
in the ES in the APBC, as discussed in the Sec.~IV C.

\section{Transition points of the bond-alternating HAF chain}

\begin{figure}[htbp]
\centering
\includegraphics[width=1.00\columnwidth]{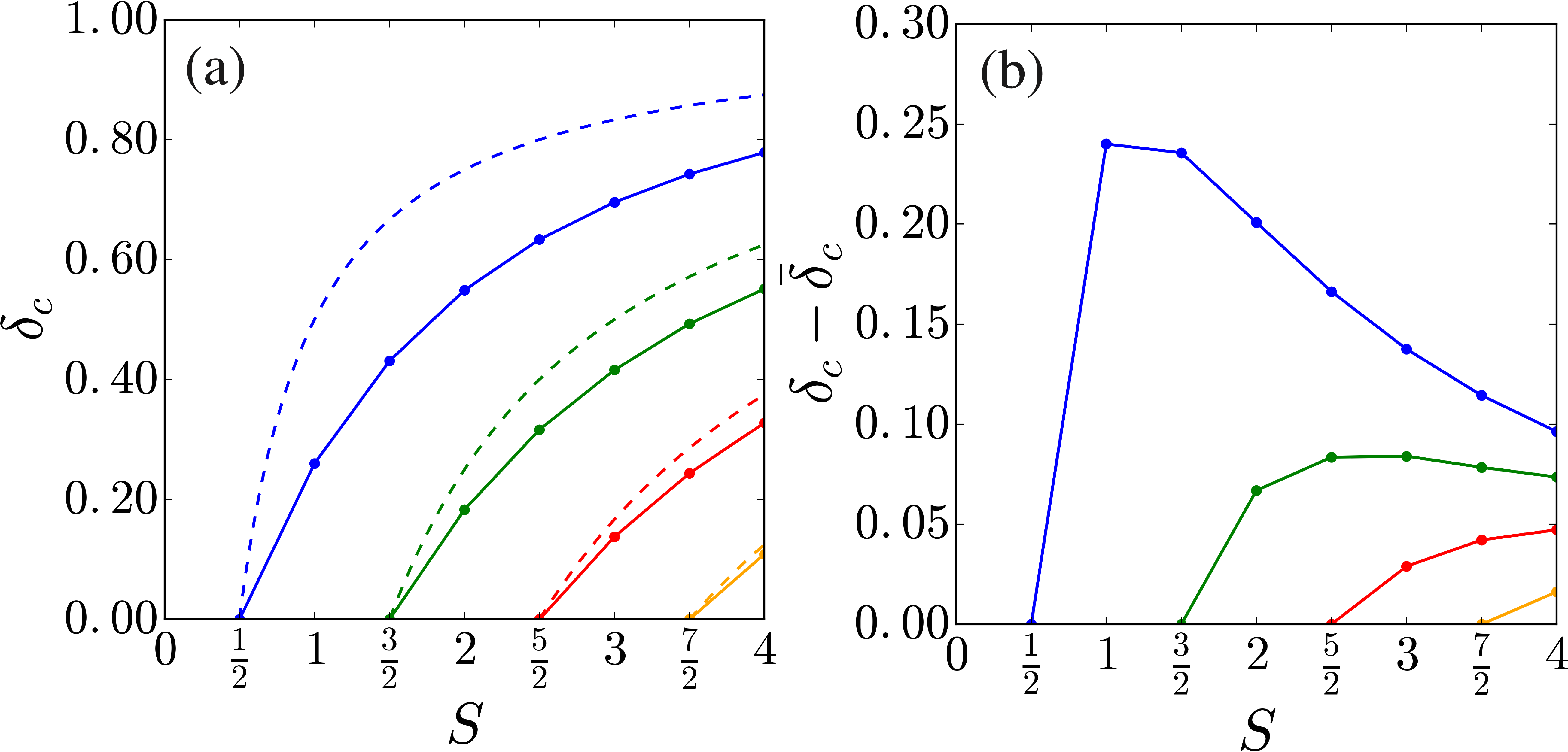}
\caption{
(Color online) (a) Transition points $\delta_c$ calculated for the bond-alternating HAF 
chain with integer and half-integer spins $S=1/2,1,\cdots,4$.  The solid and dotted lines
show the calculated results for $\delta_{c}$ in the DMRG and the results for
$\bar{\delta}_{c}$ obtained from the O(3) NLSM, respectively.  (b) The difference in
the transition points between our results and the results from the O(3) NLSM.  
}
\label{fig8}
\end{figure}

\begin{table}[htbp]
\caption{
Calculated transition points $\delta_{c}$ of the bond-alternating HAF chain with integer
and half-integer spins $S$.  The size extrapolations $L\rightarrow\infty$ are made using
the polynomial $\delta_{c}(L)=\delta_{c}(\infty)+A/L^{2}+B/L^{4}$.  
The transition points $\bar{\delta}_{c}$ obtained in the O(3) NLSM are also given.  
Our results are compared with results of the previous QMC simulation
\cite{MNakamura2002}, where the twisted order parameter is used.  
}
\centering
\begin{tabular}{|c|c|c|c|}
\hline
Transition points & $\delta_{c}$ & $\bar{\delta}_{c}$ & Previous studies \\
\hline\hline
$S=1$, (1,1)-(2,0) & 0.25995(3) & 0.500 & 0.25997(3) \\
$S=2$, (2,2)-(3,1) & 0.1831(0) & 0.250 & 0.1866(7) \\
$S=2$, (3,1)-(4,0) & 0.5491(7) & 0.750 & 0.5500(1) \\
$S=3$, (3,3)-(4,2) & 0.137(7) & 0.167 & - \\
$S=3$, (4,2)-(5,1) & 0.416(0) & 0.500 & - \\
$S=3$, (5,1)-(6,0) & 0.695(8) & 0.833 & - \\
$S=4$, (4,4)-(5,3) & 0.108(8) & 0.125 & - \\
$S=4$, (5,3)-(6,2) & 0.327(8) & 0.375 & - \\
$S=4$, (6,2)-(7,1) & 0.551(4) & 0.625 & - \\
$S=4$, (7,1)-(8,0) & 0.778(7) & 0.875 & - \\
\hline\hline
$S=3/2$, (2,1)-(3,0) & 0.4310(3) & 0.667 & 0.43131(7) \\
$S=5/2$, (3,2)-(4,1) & 0.316(5) & 0.400 & - \\
$S=5/2$, (4,1)-(5,0) & 0.633(7) & 0.800 & - \\
$S=7/2$, (4,3)-(5,2) & 0.243(6) & 0.286 & - \\
$S=7/2$, (5,2)-(6,1) & 0.492(7) & 0.571 & - \\
$S=7/2$, (6,1)-(7,0) & 0.742(7) & 0.857 & - \\
\hline
\end{tabular}
\label{table:TransitionPoints}
\end{table}

Since the discovery of the $(m,n)$-type VBS states in the bond-alternating Heisenberg chain,
the dimerization transition points $\delta_c$ have been estimated by a number of numerical 
studies using the string order parameter \cite{KTotsuka1995}, level spectroscopy
\cite{AKitazawa1997A,AKitazawa1997B,AKitazawa1997C,KNomura1998}, twisted
order parameter \cite{MNakamura2002}, and quantized Berry phase \cite{THirano2008}.  
In Fig.~\ref{fig8}, we summarize the transition points calculated for the bond-alternating HAF
chain with integer and half-integer spins $S=1/2,1,\cdots,4$, where we adopt the level spectroscopy
technique for large-size systems in the DMRG framework.  The extrapolations to $L\rightarrow\infty$
are made by the polynomial fitting of $\delta_{c}(L)=\delta_{c}(\infty)+A/L^2+B/L^4$, where we use
the systems
up to $L=80$ for $S=1$,
up to $L=46$ for $S=3/2$,
up to $L=50$ for $L=2$,
up to $L=40$ for  $S=5/2$,
up to $L=40$ for  $S=3$,
up to $L=20$ for $S=7/2$, and
up to $L=20$ for $S=4$.  
The results are given in Table \ref{table:TransitionPoints}, where we find that
our results obtained by the size extrapolation are consistent with the previous
results obtained by the QMC simulations using the twisted order parameter
\cite{MNakamura2002} but that our results are much more accurate due to
a small finite-size effect.  
As discussed in Sec.~II A, the bond-alternating HAF chain in the large-$S$
limit can be described by the O(3) NLSM [see Eq.~(\ref{eq:O3NLSM})].  
The phase transition points, in particular, can be estimated as
$\bar{\delta}_{c}=1-(2n+1)/2S$ ($n=0,1,\cdots,2S-1$), where the
$\Theta$ term becomes $\pi$.  We compare our results for the transition points
with those of the O(3) NLSM in Fig.~\ref{fig8}(b).  Our calculated results, which
converge to zero in the large-$S$ limit, are consistent with the semi-classical
treatment in the O(3) NLSM.  Moreover, the slow convergence shown in
Fig.~\ref{fig8}(b) suggests the power-law dependence of the renormalized transition
points as a function of $1/S$.  

\end{appendix}

\bibliography{reference.bib}

\end{document}